\documentclass[aps,floatfix,showpacs,twocolumn,prl]{revtex4}
\usepackage[dvips]{graphicx}
\usepackage{amsmath}
\usepackage{amsfonts}
\usepackage{amssymb}
\usepackage{times}
\usepackage{MHequ}

\def\D{{d_{\rm int}}}

\def\H{{\rm H}}
\def\C{{\rm C}}
\def\L{{\rm L}}
\def\R{{\rm R}}

\def\emu{\bar{\mu}}
\def\kB{k_{\rm B}}

\def\nablax{~\partial_x}
\def\eref#1{(\ref{#1})}
\let\rho=\varrho

\def\dof{{\em d.o.f.}}

\def\dT{\nablax\left(\frac{1}{T}\right)}

\def\dmuu{\nablax\left(-\frac{\mu}{T}\right)}

\def\B{\frac{\lambda p_t(l) l}{(2\pi  m)^{1/2}}}

\newcommand{\ave}[1]{\langle #1\rangle}

\begin{document}
\title{Increasing thermoelectric efficiency towards the Carnot limit}
\author{Giulio Casati${}^{1,2}$}
\author{Carlos~Mej\'{\i}a-Monasterio${}^3$}
\author{Toma\v z Prosen${}^4$}
\affiliation{Center for  Nonlinear and  Complex  Systems, Universit\`a
  degli Studi dell'Insubria, Como Italy${}^1$}
\affiliation{CNR-INFM and Istituto Nazionale di Fisica Nucleare, Sezione di Milano${}^2$}
\affiliation{D\'epartement de Physique Th\'eorique, Universit\'e de Gen\`eve${}^3$}
\affiliation{Physics Department,  Faculty of Mathematics  and Physics, University  of Ljubljana, Ljubljana, Slovenia${}^4$}
\date{\today}

\begin{abstract}
We study the problem of thermoelectricity and propose a simple microscopic mechanism for
the increase of thermoelectric efficiency.  We consider the cross
transport of particles and energy in open classical ergodic billiards. 
We show that, in the linear response regime, where we find exact expressions
for all transport coefficients, the thermoelectric efficiency of ideal ergodic
gases can approach Carnot efficiency for sufficiently complex charge carrier
molecules.  Our results are clearly demonstrated with a simple numerical
simulation of a Lorentz gas of particles with internal rotational degrees of
freedom.
\end{abstract}

\maketitle

Although thermoelectricity was discovered almost 200 years ago, a strong
interest of the scientific community arose only in the 1950's when Abram Ioffe
discovered that doped semiconductors exhibit relatively large thermoelectric
effect. This initiated an intense research activity in semiconductors physics
which was not motivated by microelectronics but by the Ioffe suggestion that
home refrigerators could be built with semiconductors \cite{mahan,majumdar}.
As a result of these efforts the thermoelectric material
$\textrm{Bi}_2\textrm{Te}_3$ was developed for commercial purposes.  However
this activity lasted only few years until the mid 1960's since, in spite of
all efforts and consideration of all type of semiconductors, it turned out
that thermoelectric refrigerators have still poor efficiency as compared to
compressor based refrigerators.  Nowadays Peltier refrigerators are mainly
used in situations in which reliability and quiet operation, and not the cost
and conversion efficiency, is the main concern, like equipments in medical
applications, space probes etc.  In the last decade there has been an
increasing pressure to find better thermoelectric materials with higher
efficiency.  The reason is the strong environmental concern about
chlorofluorocarbons used in most compressor-based refrigerators.  Also the
possibility to generate electric power from waste heat using thermoelectric
effect is becoming more and more interesting
\cite{dresselhaus,mahan,majumdar}.
 
The suitability of a thermoelectric material for energy conversion or
electronic refrigeration is evaluated by the thermoelectric figure of merit
$Z$,
\begin{equ} \label{eq:ZT-def}
Z = \frac{\sigma S^2}{\kappa} \ ,
\end{equ}
where $\sigma$ is the coefficient of electric conductivity, $S$ is the Seebeck
coefficient and $\kappa$ is the thermal conductivity. The Seebeck coefficient
$S$, also called thermopower, is a measure of the magnitude of an induced
thermoelectric voltage in response to a temperature difference across the
material.

For a given material, and a pair of temperatures $T_\H$ and $T_\C$ of hot and
cold thermal baths respectively, $Z$ is related to the {\em efficiency} $\eta$
of converting the heat current $J_Q$ (between the baths) into the electric
power $P$ which is generated by attaching a thermoelectric element to an
optimal Ohmic impedance. Namely, in the linear regime:
\begin{equ} \label{eq:efficiency}
\eta = \frac{P}{J_Q} = \eta_\mathrm{carnot} \cdot \frac{\sqrt{ZT + 1} -
  1}{\sqrt{ZT + 1} + 1} \ ,
\end{equ}
where $\eta_\mathrm{carnot}=1-T_\C/T_\H$ is the Carnot efficiency and $T =
(T_\H + T_\C)/2$. Thus a good thermoelectric device is characterized by a
large value of the non-dimensional figure of merit $ZT$.

Since the 1960's many materials have been investigated but the maximum value
found for $ZT$ was achieved for the
$(\textrm{Bi}_{1-x}\textrm{Sb}_x)_2(\textrm{Se}_{1-y}\textrm{Te}_y)_3$ alloy
family with $ZT \approx 1$.  However, values $ZT > 3$ are considered to be
essential for thermoelectrics to compete in efficiency with mechanical power
generation and refrigeration at room temperatures.  The efforts recently
focused on a bulk of new advanced thermoelectric materials and on
low-dimensional materials, and only a small increment of the efficiency, $ZT
\lesssim 2.6$, has been obtained \cite{dresselhaus}.

One of the main reasons for this partial success is a limited understanding of
the possible microscopic mechanisms leading  to the increase of $ZT$, with few
exceptions  \cite{linke}. From  a dynamical  point of  view, cross  effects in
transport  have  been barely  studied  \cite{microscopic,MMLL}.   So far,  the
challenge lies in engineering a material for which the values of $S$, $\sigma$
and  $\kappa$  can  be   controlled  independently.   However,  the  different
transport  coefficients  are  interdependent,  making  optimization  extremely
difficult.

In this  paper we take  a completely different approach.
Inspired by  kinetic theory  of ergodic gases  and chaotic billiards,  we show
that large  values of $ZT$,  in principle approaching to  Carnot's efficiency,
can be  obtained when the energy of  the carrier particles does  not depend on
the thermodynamic forces.

In the linear response regime (see {\em e.g.}  \cite{bergman}), one writes the
general expressions for the heat current $J_Q$ and the electric current $J_e$
through an homogeneous sample subjected to a temperature gradient $\nablax T$
and a electrochemical potential gradient $\nablax \emu$ as
\begin{equa}[2] \label{eq:Ju1}
& J_Q & \ = \ & - \kappa' \nablax T - T \sigma S \nablax\emu\ ,
\\
& J_e & \ = \ & - \sigma S \nablax T - \sigma \nablax\emu\ .
\end{equa}
Here  and in  what follows,  we  assume that  the transport  occurs along  the
$x$-direction  and the  temperature  is  given in  units  where the  Boltzmann
constant $\kB=1$.

The electrochemical  potential is the sum  of a chemical and  an electric part
$\emu = \mu  + \mu_e$, where $\mu$ is the chemical  potential of the particles
and, if $e$ is the particle's charge,  $\mu_e = e\phi$ is the work done by the
particles against an external electric field ${\cal E}=-\nablax\phi$.
From (\ref{eq:Ju1}) the usual phenomenological relations follow: if the
thermal gradient vanishes, $\nablax T=0$, then $J_e = -\sigma \nablax \phi =
\sigma {\cal E}$, since for an isothermal homogeneous system $\mu$ is uniform.
If the electric current vanishes, $J_e = 0$, then $\nablax\emu = S \nablax T$,
which is the definition of the Seebeck coefficient, and $J_Q = -\kappa \nablax
T$ where $\kappa = \kappa' - T \sigma S^2$ is the usual thermal conductivity
(see {\em e.g.} \cite{domenicali}).

From the theory of irreversible thermodynamics,  $\mu$ and $\mu_e$ cannot be 
determined separately;  only  their  combination  in  $\emu$  appears  in  \eref{eq:Ju1}
\cite{walstrom}.  Based on  this equivalence,  in  what follows  we take  into
account the chemical part only, {\em i.e.}, $\bar{\mu} = \mu$.

Our aim is  to study thermoelectricity from  an ``energy transport'' point
of view. To linear order, the energy and particle density currents ${J}_u$ and
${J}_\rho$  respectively,  can be  written  in  terms  of the  Onsager  matrix
$\mathbb{L}$ \cite{domenicali,callen} as
\begin{equa}[2] \label{eq:Ju2}
& J_u & \ = \ & L_{uu}\dT + L_{u\rho}\dmuu , \\ 
& J_\rho& \ = \ & L_{\rho u}\dT + L_{\rho\rho}\dmuu\ ,
\end{equa}
where  $J_e=e  J_\rho$.   In  the  absence of  magnetic  fields,  the  Onsager
reciprocity  relations states  that $\mathbb{L}$  is symmetric,  $\L_{u\rho} =
L_{\rho u}$. From the entropy balance equation for open systems
\begin{equ} \label{eq:Ju-def}
J_u = J_Q + \mu J_\rho \ ,
\end{equ}
and substituting  $J_Q$ in \eref{eq:Ju1} in  favor of $J_u$  and comparing the
resulting  equations   with  \eref{eq:Ju2}  it  follows   that  the  transport
coefficients can be written in terms of the $L$-coefficients as
\begin{equ}\label{eq:coefs}
\sigma  =  \frac{e^2}{T}L_{\rho\rho} \ , \quad
\kappa  =  \frac{1}{T^2}\frac{\det \mathbb{L}}{L_{\rho \rho}} \ , \quad
S  =  \frac{1}{eT}\left(\frac{L_{u\rho}}{L_{\rho\rho}} - \mu\right) \ .
\end{equ}
Eqs.~\eref{eq:Ju2}  and  \eref{eq:coefs}  are  completely  equivalent  to  the
description \eref{eq:Ju1}.  Furthermore,  from Eq. \eref{eq:ZT-def}, we obtain
for the figure of merit
\begin{equ}\label{eq:ZT}
ZT = \frac{\left(L_{u\rho} - \mu L_{\rho\rho}\right)^2}{\det \mathbb{L}}\ .
\end{equ}
Expressions \eref{eq:Ju2} and \eref{eq:ZT} provide a very convenient way for
numerical or analytical evaluation of $ZT$ for different kinds of dynamical
models.

The second law of thermodynamics only requires that $\mathbb{L}$ is positive
definite. Therefore, from \eref{eq:ZT} it is clear that the second law does
not impose any upper bound on the value of $ZT$.  Furthermore, the crucial
observation is that the Carnot's limit $ZT=\infty$ is reached when the energy
density current and the electric current become proportional, since then $\det
\mathbb{L}= 0$.  Suppose for example that both energy and charge are carried
only by non-interacting particles, like in a dilute gas.  Then the microscopic
instantaneous currents per particle at position $x^*$ and time $t$, are
\begin{equa}[2] \label{eq:inst-currents}
& j_u(x^*,t) & \ = \ & E(t) v_x(x(t),t)\delta(x^*-x(t)) \ , \\
& j_e(x^*,t) & \ = \ & e v_x\delta(x^*-x(t)) \ ,
\end{equa}
where $E$ is the energy of the particle, $x$ its position and $v_x$ its
velocity along the field.  The thermodynamic averages of the two currents
(appearing in Eq.~\eref{eq:Ju2}) become proportional precisely when the
variables $E$ and $v_x$ are {\em un-correlated}
\begin{equ} \label{eq:main}
J_u = \ave{j_u} =  \ave{E}\ave{v_x} =  \frac{\ave{E}}{e}\ave{j_e}  = \ave{E}J_\rho\ .
\end{equ}
Therefore,  $ZT=\infty$ follows  from  the fact  that  the average  particle's
energy $\ave{E}$ does not depend  on the thermodynamic forces.  In the context
of classical physics this happens for instance in the limit of large number of
internal degrees  of freedom (\dof),  provided the dynamics is  {\em ergodic}.

\begin{figure}[!t]
\begin{center}
  \includegraphics[scale=0.36]{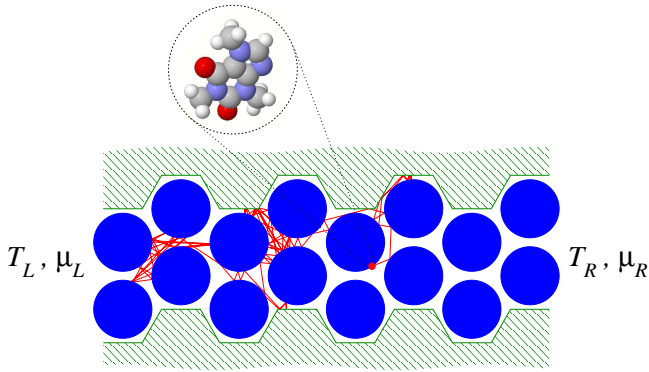}
\caption{ The open Lorentz gas system and a typical particle's trajectory.
The composite particle (schematically represented as a molecule) is scattered
from fixed disks of radius $R$ disposed in a triangular lattice at critical
horizon, {\em i.e.}, the width and height of the cells are $\Delta x = 2R$ and
$\Delta y = 2W$ respectively, where $W=4R/\sqrt{3}$ is the separation between
the centers of the disks. The channel is coupled at the left and right
boundaries to two thermochemical baths at temperatures $T_\L$ and $T_\R$ and
chemical potentials $\mu_\L$ and $\mu_\R$, respectively.\label{fig:model-1}}
\end{center}
\end{figure}

We consider an ergodic  gas of non-interacting, electrically neutral particles
of mass  $m$ with  $d_{\rm int}$ internal  \dof ~(rotational  or vibrational),
enclosed in a  $d$ dimensional container.  To study  the non-equilibrium state
of  such  dilute  poly-atomic  gas  we consider  a  chaotic  billiard  channel
(Fig.~\ref{fig:model-1}) connected  through openings of size  $\lambda$ to two
reservoirs of particles which are idealized as infinite chambers with the same
poly-atomic gas at  equilibrium density $\rho$ and temperature  $T$.  From the
reservoirs,  particles  are injected  into  the  channel  at a  rate  $\gamma$
obtained by integration over  energy of the appropriate canonical distribution
to give
\begin{equ} \label{eq:gamma}
\gamma \ = \ \frac{\lambda}{\left(2\pi m\right)^{1/2}}\rho T^{1/2} \ .
\end{equ}

The particle injection  rate $\gamma$ is related to the  value of the chemical
potential $\mu$ at the reservoirs which, for polyatomic molecules with a total
of $D=d + d_{\rm int}$ \dof, reads
\begin{equ} \label{eq:chempotMM}
\mu = T \ln\left(\frac{c_D \ \gamma}{T^{(D+1)/2}}\right) \ , 
\end{equ}
where $c_D$ is a $D$-depending constant.  Furthermore, averaging the energy of
the {\em injected} particles over the canonical distribution, denoted as $\ave{E}$, we obtain
the rate at which energy is injected from the reservoirs as $\varepsilon =
\gamma \ave{E} = \gamma T(D+1)/2$.

Let $p_t(l)$ be the transmission probability of the channel of length $l$. For
a billiard system of noninteracting particles \cite{phonons}, the density
currents $J_u, J_\rho$ assume a simple form: they are $p_t(l)$ times the
difference between the left and right corresponding injection rates,
$\varepsilon,\gamma$, respectively, namely
\begin{equ} \label{eq:noint-currents-1}
J_\rho \ = \  p_t\left(\gamma_\L - \gamma_\R\right)\ , \quad \quad
J_u    \ = \  p_t\left(\varepsilon_\L -  \varepsilon_\R\right)\ .
\end{equ}

Using \eref{eq:chempotMM} to eliminate $\gamma$ in favor of $\mu$ we obtain,
\begin{equa}[2] \label{eq:noint-currents-2}
&J_\rho &\ = \ &  -\B ~\nablax\left(T^{(D+1)/2} \ e^{\ \mu/T}\right) \
,\\  
&J_u & \ = \ & -\B ~\frac{D+1}{2}\nablax\left(T^{(D+3)/2} \ e^{\ \mu/T}\right)\ ,
\end{equa}
Taking total differentials of \eref{eq:noint-currents-2} in the variables
$1/T$ and $\mu/T$ and comparing the resulting expression with
Eq.~\eref{eq:Ju2} we obtain exact microscopic expressions for the Onsager
coefficients, namely
\begin{equa}[2] \label{eq:L-coeffs}
& L_{\rho\rho} & \ = \ & \B \rho T^{1/2} \ , \\
& L_{\rho u} = L_{u\rho} & \ = \ & \B\left(\frac{D+1}{2}\right) \rho
T^{3/2} \ , \\ 
& L_{uu} & \ = \ & \B \frac{(D+1)(D+3)}{4} \rho T^{5/2} \ .
\end{equa}
Note that for a chaotic billiard channel with a diffusive dynamics, the
transmission probability decays as $p_t(l) \propto l^{-1}$ which means that
all the elements of the Onsager matrix $\mathbb{L}$ become size independent.

Finally, plugging \eref{eq:L-coeffs} into \eref{eq:ZT} and noting that
$c^*_V=D/2$ is the dimensionless heat capacity at constant volume of the gas,
we obtain
\begin{equ}\label{eq:ZTmain}
ZT = \frac{1}{\hat{c}_V}\left(\hat{c}_V - \frac{\mu}{T}\right)^2 \ ,
\end{equ}
where for simplicity we have called $\hat{c}_V = c^*_V + 1/2$. A particular
case of \eref{eq:ZTmain} was previously obtained, for noninteracting
monoatomic ideal gases in $3$ dimensions \cite{vining}.

In absence of  particles' interaction, $ZT$ is independent  of the sample size
$l$  and  depends on  the  temperature  only  through the  chemical  potential
term. This is due to the fact  that with no interactions, $p_t$ depends on the
geometry of the billiard only.  From  a physical point of view this means that
the mean free path of the gas particles is energy independent.  Were particles
interacting, $p_t$  would depend on the  local density and  temperature of the
gas, leading to a  more realistic situation \cite{EMMZ}. Interestingly enough,
we have found  that Eq.~\eref{eq:ZTmain} is an upper  limit of the interacting
case when the interaction strength vanishes \cite{ztlong}.

We  shall  now  confirm  Eq.~\eref{eq:ZTmain}  with a  very  simple  numerical
demonstration of  a $2$-dimensional chaotic  Lorentz gas channel  of particles
elastically colliding with circular obstacles  of radius $R$. In what follows,
we fix the  unit length setting $R=1$.  The geometry of the model  is shown in
Fig.\ref{fig:model-1}.

We  consider composite  particles with  $\D \ge  1$ internal  rotational \dof.
Each ``particle" of mass $m$ can be imagined as a stack of $d_{\rm int}$ small
identical  disks of  mass $m/\D$  and radius  $r \ll  R$, rotating  freely and
independently at a constant  angular velocity $\omega_i$, $i=1,\ldots\D$.  The
center of mass of the particle moves with velocity $\vec{v}=(v_x,v_y)$.

At each  collision of the particle  with the boundary of  the billiard (either
one of the circular obstacles or  the outer wall) an energy exchange among all
the $D$ \dof~ occurs according to the following collision rules
\begin{equa}[2] \label{eq:colrules}
& v_n' & \ = \ & -v_n \ ,\\
& v_t' & \ = \ & \frac{1-\eta\D}{1+\eta\D} v_t +
\frac{2\eta}{1+\eta\D}\sum_{k=1}^\D \omega_k \ ,\\
& \omega_i' & \ = \ & \frac{2}{1+\eta\D} v_t \sum_{k=1}^\D\left(\delta_{ik} -
  \frac{2}{\D(1+\eta\D)}\right) \omega_k \ ,
\end{equa}
where  $(v_n,v_t)$ are  the  normal  and tangent  components  of the  particle
velocity  at the collision  point and  the parameter  $\eta=\Theta/mr^2$, with
$\Theta$ being the  moment of inertia of each small fictitious internal disk.
The  primed (unprimed)  quantities refer  to their  values after  (before) the
collision. These collision  rules are a generalization of  the ones introduced
in \cite{MMLL}. Thus, they are deterministic, time reversible and preserve the
energy and local angular momentum.  The derivation of the collision rules will
appear elsewhere \cite{ztlong}.

First we considered a closed system in a finite container and we checked
energy equipartition among all \dof.  Then we have opened the system from both
ends and allowed it to exchange particles with the two baths at temperatures
$T_{\rm L}$, $T_{\rm R}$ and with chemical potentials $\mu_{\rm L}$, $\mu_{\rm
R}$. The coupling among the system and the baths is defined as follows:
whenever a particle in the system crosses the opening which separates it from
the bath, it is removed from the system.  On the other hand, with frequency
$\gamma$, particles are injected into the system, with a velocity distributed
according to the canonical distribution at the corresponding temperature.
\begin{equa}[2] \label{eq:dist}
&P_n(v_n) & \ = \ & \ \frac{m}{ T}|v_n|
\exp\left(-\frac{m v_n^2}{2 T}\right) \ ,\\
&P_t(v_t) & \ = \ & \sqrt{\frac{m}{2\pi  T}}
\exp\left(-\frac{m v_t^2}{2 T}\right) \ ,\\
&P(\omega_i) & \ = \ & \sqrt{\frac{m}{4D\pi  T}}
\exp\left(-\frac{m \omega_i^2}{4D T}\right) \ ,
\end{equa}
reflecting the assumption that the bath is an ideal gas at equilibrium
temperature $T$.

\begin{figure}[!t]
\begin{center}
  \includegraphics[scale=0.7]{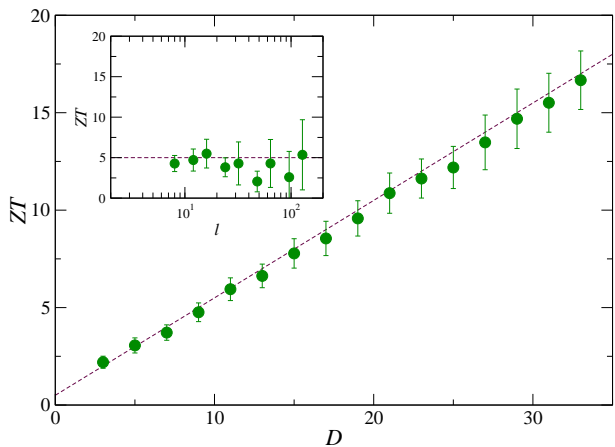}
\caption{Figure of merit $ZT$  as a function of the number of  \dof, at $\mu =
0$.  For each $D$, the transport coefficients were obtained from two different
simulations in a channel of  $10$ cells, at mean particle density $\rho=0.446$
and mean temperature $T=1000$ with: $a$) $\Delta T/T = 0.04$, $\Delta(\mu/T) =
0$,  and $b$)  $\Delta  T  = 0$,  $\Delta(\mu/T)=0.04$.   The injection  rates
$\gamma_L$   and   $\gamma_R$  are   obtained   from  $\Delta(\mu/T)$,   using
\eref{eq:chempotMM}  and  \eref{eq:gamma}.   The  dashed line  corresponds  to
$(D+1)/2$. In the  inset, the dependence of $ZT$ on the  length of the channel
$l$  is shown for  $D=9$.  The  dashed line  shows the  corresponding expected
value.
\label{fig:ZT}}
\end{center}
\end{figure}

By  two   different  simulations  with   two  linearly  independent   sets  of
thermodynamic  forces  we have  numerically  determined  the Onsager's  matrix
$\mathbb{L}$ and the value  of $ZT$ (see Fig.\ref{fig:ZT}).  Numerical results
excellently  confirm the  theoretical prediction  (\ref{eq:ZTmain}).   We have
also carefully checked that all Onsager coefficients, or conductivities, decay
with  the  size  $l$ of  the  system  as  $1/l$  which indicates  a  diffusive
transport.

\begin{figure}[!t]
\begin{center}
  \includegraphics[scale=0.7]{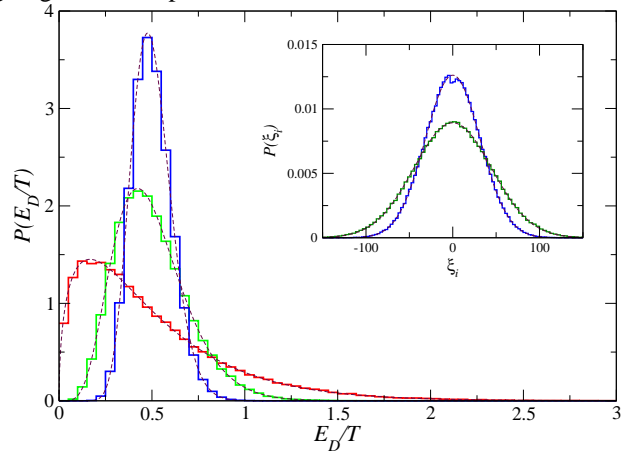}
\caption{  Probability  distribution function  of  the  energy  per degree  of
freedom  $E_D  = E/D$  determined  from  equilibrium  simulation with  $T_{\rm
L}=T_{\rm R},\mu_{\rm L}=\mu_{\rm R}$, for different number of freedoms: $D=3$
(red),  $D=13$  (green)  and  $D=43$   (blue).   The  dashed  curves  are  the
theoretical  (``Chi-square"  $\chi^2_{D}$)  distributions  of $E_D$.   In  the
inset,  the  corresponding probability  distribution  functions  for the  $x$-
component of the velocity $P(v_x)$ (blue)  and for the angular momentum of one
of the particle's disks $P(\omega_i)$ (green) is shown, for $D=3$.  The dashed
curves are the theoretical Gaussian distributions.
\label{fig:PE}}
\end{center}
\end{figure}

The  simple mechanism  for the  growth of  $ZT$ with  $d_{\rm int}$  is nicely
illustrated in  Fig.\ref{fig:PE} which shows that the  particle velocity $v_x$
has  a  Maxwellian  (Gaussian)  distribution (inset),  while  the  equilibrium
distribution of the  particle energy per degree of  freedom $E_D$ becomes more
and more  sharply peaked, and thus  {\em de-correlated} from  $v_x$ as $d_{\rm
int}$ grows.

In conclusion, we have discovered a simple general theoretical mechanism which
may find a  way to applications of thermoelectricity  in real world materials.
Even though the case of an ionized polyatomic gas may seem a little artificial
in  this context, there  may be  other important  instances where  each charge
would  be carried  by {\em  many} effectively  classical \dof.   We  have also
performed  the   first  numerical  computation  of   $ZT$  from  deterministic
microscopic equations of motion. Our method can easily be implemented for more
realistic models where also quantum effects can be taken into account.
 
The  authors  are  indebted to  H.   Linke  and  C.  Vining  for  enlightening
discussions  and correspondence,  and thank  the hospitality  of  the Institut
Henri Poincar\' e,  Paris, where part of this work  was done.  TP acknowledges
support  from grants  P1-0044 and  J1-7347 of  Slovenian research  agency. CMM
acknowledges  support  from  Fonds  National  Suisse  and  a  Lagrange
fellowship from Fondazione CRT.


\begin{thebibliography}{1}
\bibitem{phonons} Assuming that particles do not interact means that we fully
  neglect the phonon transport.
\end{thebibliography}
\end{document}